\documentclass[trackchanges,twocolumn]{aastex701}

\usepackage{amssymb}
\usepackage{lipsum}
\usepackage{longtable}
\usepackage{booktabs}
\usepackage{amsmath}
\usepackage{url}

\usepackage{xcolor}
\usepackage{mathptmx}
\begin{document}

\title{On extracting Science information from a large data set of RXTE X-ray binary spectra using KDE technique}

\author{Ranjeev Misra}
\affiliation{Inter-University Centre for Astronomy and Astrophysics (IUCAA), Pune, Maharashtra 411007, India}
\email[show]{rmisra@iucaa.in}

\author{Akash Garg}
\affiliation{Inter-University Centre for Astronomy and Astrophysics (IUCAA), Pune, Maharashtra 411007, India}
\affiliation{Persistent Systems Ltd., Pune, Maharashtra 411006, India}
\email{akash.garg@iucaa.in}

\author{Aman Kumar}
\affiliation{Inter-University Centre for Astronomy and Astrophysics (IUCAA), Pune, Maharashtra 411007, India}
\affiliation{Department of Physics, Savitribai Phule Pune University, Ganeshkhind, 411007, Pune, India}
\email{kumaramanasci@gmail.com}

\author{Ajit Kembhavi}
\affiliation{Inter-University Centre for Astronomy and Astrophysics (IUCAA), Pune, Maharashtra 411007, India}
\email{akk@iucaa.in}

\begin{abstract}

 Machine Learning/Neural Networks have been successively and extensively used in Astronomy (and other subjects) for class identification of objects based on large data sets. However, it remains difficult to extract other science information, such as what the characteristics of the class are that allow for definite identification, and whether there are objects in the sample that are unique and hence of possible science interest. An alternate approach based on rank ordering of variables and Kernel Density Estimation (KDE) may be used for class identification and to infer such science information. We test the method on a previously analyzed data set of RXTE spectra of low-mass X-ray binaries and find that black hole and neutron star systems can be distinguished at an accuracy level of $\sim 93\%$, which is comparable to deep learning techniques. The soft state spectra of black hole systems are distinct from neutron star ones. For the whole sample, observations of the pulsar 4U 1907+097 (which was inadvertently introduced into the data set) were clearly found to be unique. The previously reported hyper-soft states of  GRO J1655-40 and XTE J1550-564 are found to be unique among black hole spectra. For the rest of the neutron stars, the method identifies some unreported spectra of EXO 1745-248 as unique because they seem to be relativistic reflection-dominated spectra. We discuss the potential and scalability of this method. 
 
 \end{abstract}



\section{Introduction} 

Over the last few decades, Neural Networks/Machine Learning (NNML) have played a critical role in Astrophysics in class identification using large data sets for training. More recently, the generative ability of modified techniques to predict next elements in a series has led to the advent of Artificial Intelligence and, in particular, Large Language Models. This ability has been directly exploited in Astrophysics to predict the temporal behavior of systems \citep[e.g., Solar foundation model][]{roy2025}. Nevertheless, class identification remains an important utility of these techniques, especially for analysis of large data sets.

A drawback of these models is that it is difficult to interpret the results obtained and to infer other scientific information from the data sets. A generic feature of these techniques is the tuning of weights to minimize a loss function, and it is difficult to infer physical meanings from the weights. Thus, the trained system has to be often looked upon as a 'black box' where inferences can be obtained by examining the outputs for particular inputs. The importance of an input variable can be obtained by eliminating that variable or by estimating the gradient of the predicted probability with respect to the variable value. However, this does not provide information regarding any non-linear correlation between the different variables that allow the system to predict a class.  One could also consider the set of input variables that produces the highest probability value and ascribe the set as a characteristic trait for that class. However, that would not necessarily be correct, since the system considers an object to be of a certain class if the probability is higher than those of other classes, e.g., $> 50\%$ when there are two classes. So a higher value of probability does not contribute to the loss function, and hence probabilities much greater than $50\%$ may not have meaning. Moreover, the tuned weights depend on the initial state of the system before training, and hence one can get different probability results for the same training data sets. Thus, these factors make scientific interpretations difficult.

Kernel Density Estimation (KDE), also known as Parzen Density Window Estimation, has been used to obtain a smooth density function from sparse data as well as for classification and class identification \citep[e.g.,][]{taylor97}. It has been used for a wide variety of applications, such as to estimate galaxy densities \citep{ferdosi11}, flux distribution of radio sources \citep{liu2025}, quasar photometric redshifts \citep{yang2017}, and to produce Gamma-ray sky maps \citep{holler2024}. It has been used to distinguish different types of supernovae \citep{Newling2011} and star clusters \citep{nambiar2019}.

In this work, we describe a technique called the Direct Incidence Method (DIM), which is based on a modified KDE using rank-ordered variable values. We apply the method to a previously analyzed data set of RXTE X-ray binary spectra \citep{pattnaik,garg26} to distinguish black holes and neutron star systems and to find unique spectra in the set. The next section describes the technique, while the appendix discusses its possible scaling and generalization. Section 3.1 introduces the data set and lists the reasons for choosing the data set. The remaining three subsections of Section 3 present the results of the analysis, while the last section discusses the results.

\section{Direct Incidence Method} \label{sec:style}

We consider a large data set consisting of $N_{rows}$  rows and $N_{var}$ number of columns of real values, which are labelled as $V_1, V_2$...$V_{N_{var}}$. There is also a corresponding column denoting the class of each row, where the class is denoted as an integer that ranges from $c = 0, 1, \dots, c_{max}$. We discuss generalization of the data set to different formats in Appendix \ref{scale}.

Each of the variables, $V_j$, is first individually rank-ordered, and the rank ordering is binned into $N_{bin}$, and each of the bins is labeled with an integer $IV_j = 0,1, \dots, N_{bin}- 1$. Then each real value of $V_j$ is replaced by its corresponding rank order integer $IV_j$. Thus if $N_{bin}$ is ten, then the real value of $V_j$ is replaced by $9$ if the value is greater than 90 percentile and by $8$ if it lies between 80 and 90 percentile and so on. In this work we consider $N_{bin}$ to be a parameter of the analysis and take it to be the same for all variables, $j = 1, \dots ,N_{var}$.

Every row $i$ can then be denoted by a set of integers $IV^i_j$, which for convenience we denote as a vector $\vec{IV^i}$ of dimension $N_{var}$. Note that the components of the vectors are all integers. The total number of possible vectors is finite and equal to $N_{bin}^{N_{var}}$. The incidence of a class $c$ as a function of $\vec{IV}$, $I_c(\vec{IV})$, is the number of rows that correspond to the vector $\vec{IV}$ and the class $c$. In other words, the incidence is obtained by counting the number of times the vector $\vec{IV}$ appears in the data set and has the corresponding class value of $c$. 

In KDE, the Possibility of a vector being associated with a class, $c$, $P_c(\vec{IV})$ is obtained from the $I_c$ of all vectors weighted by the distance between the vectors. Thus,

\begin{equation}
  P_c(\vec{IV}) =\frac{1}{N_c} \sum_{\vec{IV^\prime}} I_c(\vec{IV^\prime})/(D+s)^p
\end{equation}

where $N_c$ is the total number of rows in the data set that correspond to class $c$ and the summation is over all possible vectors $\vec{IV}^\prime$. The distance $D$ is the Euclidean distance between vectors $\vec{IV}^\prime$ and  $\vec{IV}$ i.e.,

\begin{equation}
  D = \sqrt{\sum_{j=1}^{N_{var}} \left( IV^\prime_j - IV_j \right)^2}
  \label{eq2}
\end{equation}

Typically, the kernel used in KDE is of a Gaussian form. However, here we use a power-law profile, with index $p$ and the softness parameter $s$ as two more parameters of the analysis. The softness parameter is there to ensure that $P_c(\vec{IV})$ does not blow up when $D = 0$, i.e., when $\vec{IV^\prime} = \vec{IV}$. We have tested that this form produces similar results to the Gaussian Kernel, but has some implementation advantages that we discuss in the next section.

The Possibilities, $P_c(\vec{IV})$, can then be normalized to give a probability that a vector $\vec{IV}$ corresponds to a class $c$, by

\begin{equation}
  p_c(\vec{IV}) = \frac{P_c(\vec{IV})}{\sum_{j=0}^{j=c_{max}} P_j(\vec{IV})}
  \label{prob}
\end{equation}

Another useful quantity is the Uniqueness as a function of vectors, which can be defined as

\begin{equation}
  U (\vec{IV}) =  {\sum_{j=0}^{j=c_{max}} \frac{1}{N_c} \sum_{\vec{IV^\prime}} I_c(\vec{IV^\prime}) D^{pd}}
  \label{unique}
\end{equation}

which measures the weighted distance between the vector $\vec{IV}$ and others, and the index $pd$ is another analysis parameter.

\begin{figure*}
    \centering
    \includegraphics[width=0.48\linewidth]{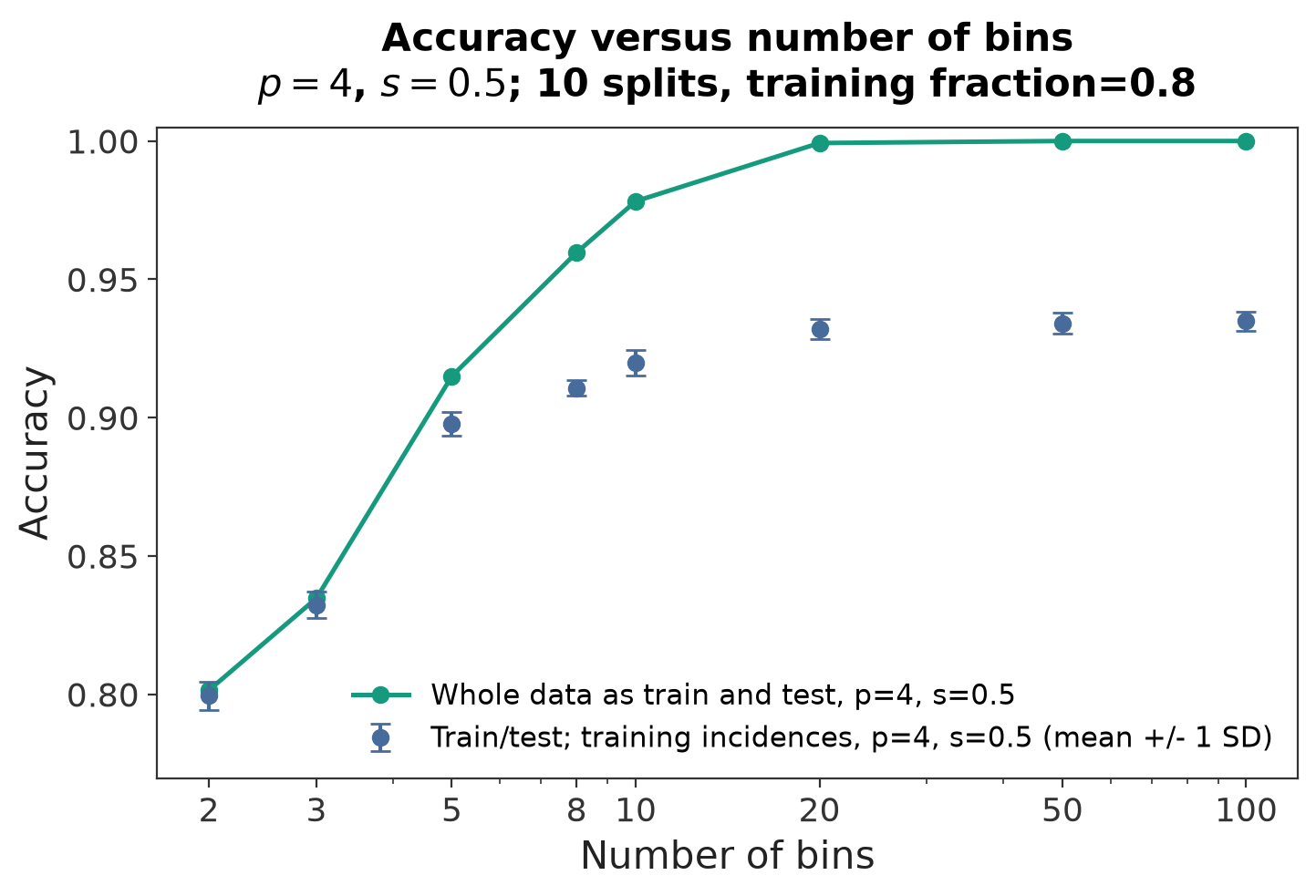}
    \includegraphics[width=0.48\linewidth]{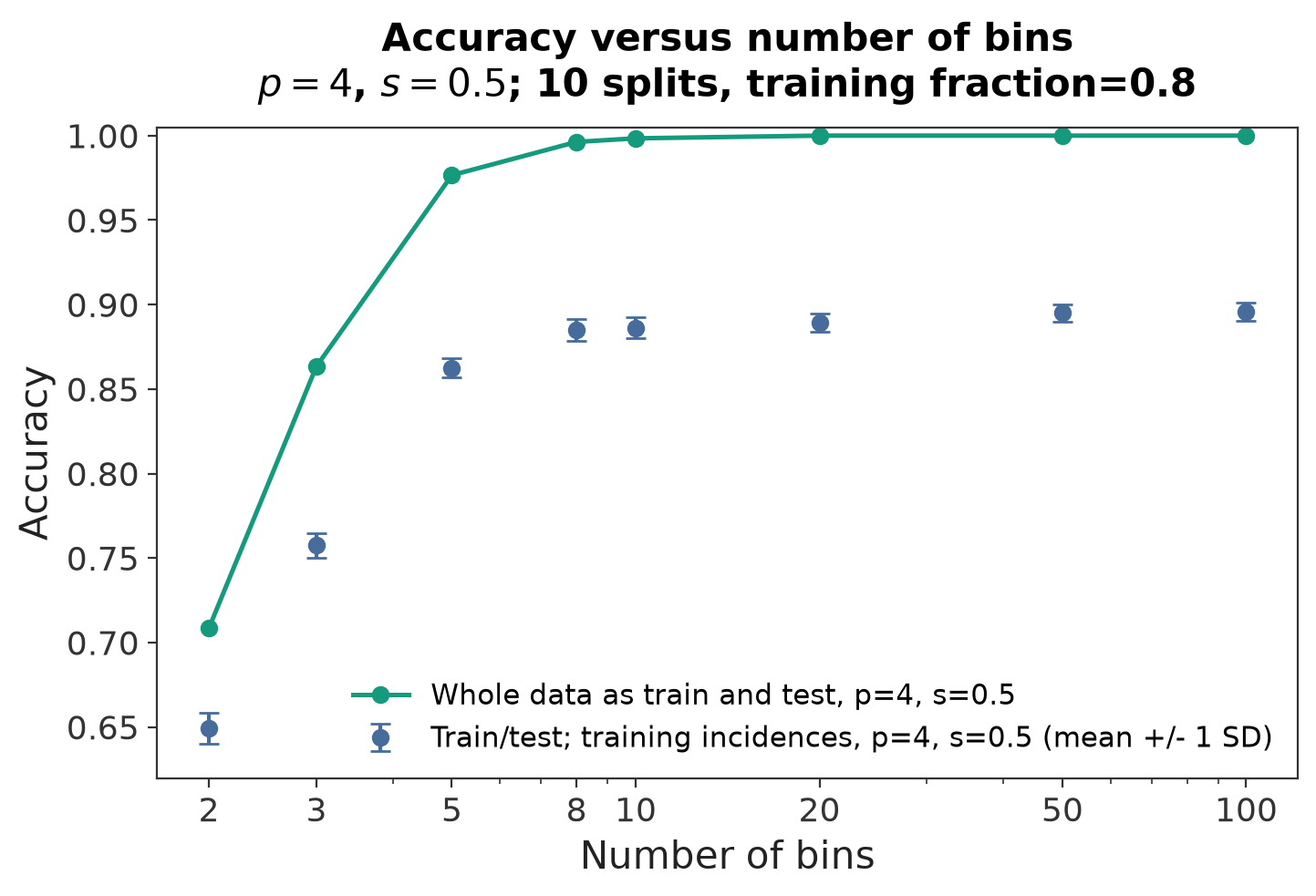}
    \caption{Classification accuracy as a function of the number of bins, $N_{\rm bin}$, for $s = 0.5$ and $p = 4$. \textit{Left:} the five derived spectral parameters. \textit{Right:} the raw count spectra of 43 spectral bins, shown for illustration. The mean accuracy over ten random 80\%/20\% splits of the observations into base and test sets is plotted, with the standard deviation as the uncertainty. The solid line shows the accuracy when the full data set is used as both the base and test sets.}
    \label{fig:accuracy_bins}
\end{figure*}

The probabilities and the Uniqueness are derived from the Incidences, $I_c(\vec{IV})$, which need to be obtained and stored from the original data. The Incidences are integers, and the total number of vectors is $N_V = N_{bin}^{N_{var}}$, which can be very large. However, one needs to store only non-zero values of the Incidences, i.e., only for the number of vectors with non-zero Incidences, $N_{IN0}$. $N_{IN0}$ can be substantially smaller than $N_V$ and cannot be larger than the total number of rows in the data set $N_{rows}$. In practice, one can define a two-dimensional integer array $IA(c,k) = I_c(\vec{IV})$ where $k = 1, \dots, N_{IN0}$ and another long integer array $Index(k)$. $Index(k)$ would then have to uniquely define $\vec{IV}$ and this can be done by

\begin{equation}
  Index(k) = \sum_{j=1}^{j=N_{var}}IV_j N_{bin}^{j-1}
\end{equation}

Note that since $IV_j < N_{bin}$, $Index(k)$ can be used to generate the specific corresponding $IV_j$. If $N_{bin} = 10$, then $Index (k)$ would be the decimal notation for $IV_j$. Since the biggest long integer which can be defined in most computer languages is $2^{16}-1$, this limits $N_V = N_{bin}^{N_{var}}$ to be less than this value. We discuss possible methods to overcome this limitation in Appendix \ref{scale}.

The method can be applied to do the following:-

$\bullet$ Given a new object with variables, it can predict the probability of the object belonging to a certain class using the probabilities $p_c(\vec{IV})$. The accuracy of such predictions can be tested by dividing the data set into two groups, one for training (i.e., computation of the Incidences)  and the other for testing. Multiple realizations of the division into two groups can provide a standard deviation of the accuracy. An important aspect of such an analysis would be to optimize the analysis parameters, which are $N_{bin}$, $p$, and $s$, to obtain the highest accuracy.

$\bullet$ For the whole data set, one can compute the number of vectors (as a fraction of the total number of vectors, $N_V$) that have a probability between $p_c$ and $p_c+\delta p_c$, thereby producing a distribution function of the probabilities. This not only provides an overall picture of the classification scheme, but can also isolate those vectors which yield the highest probabilities for a certain class. Thus, one could then infer that if an object corresponds to this vector (and hence certain variable ranges), then it is most likely to be a member of a certain class. In other words, it may give the characteristic variable values that identify the object to be of that class.

$\bullet$ One can compute Uniqueness for all vectors that are in the data set (i.e., those vectors which have at least one non-zero Incidence). The vectors with the highest value of Uniqueness will then represent the isolated or unique variable space. The specific rows in the data set (i.e., the objects) which have these high uniqueness value vectors will correspond to unique objects. These can then be subjected to further analysis to understand why they are unique. Such an exercise can be undertaken for the whole data set or by redoing the analysis for only one particular class, which would then provide unique objects of that class.

\section{Analysis of RXTE spectra of low-mass X-ray binaries}

\subsection{The Data Set}

To test the scheme, we use the same data set as used by \cite{pattnaik} and \cite{garg26}  to classify black holes and neutron star low-mass X-ray binaries using a large sample of RXTE spectra.

There are several reasons why we have chosen this particular data set to test the method. The data is in the format mentioned in Section 2, which is that it consists of a set of real variables (spectral counts or parameters) and an integer classification (0$-$ Neutron star, 1$-$ Black Hole). The source selection, data reduction, and preparation of this publicly available data have already been done. The data set has already been subjected to different neural network classification schemes, and it has been shown that the random forest algorithm provides $\sim 87\%$ classification accuracy while a deep learning method yields $\sim 93\%$, which allows us to benchmark the present method. \cite{garg26} have also used spectral parameters (instead of the raw spectral counts data), which allows for more direct extraction of science information. The RXTE spectra of these sources have been extensively studied over the last several decades, which allows us to assess the results obtained in this work.

The data consist of $14855$ observations from 33 neutron star systems and 28 black hole systems. \cite{pattnaik} use 43-channel raw data counts (roughly corresponding to the energy range 5-25 keV), which are Crab-normalized to take into account the effective area and gain changes of the instrument. \cite{garg26} analyzed the same data set using deep learning neural networks. In addition, they fitted all observations with a power-law and blackbody model and used the resultant spectral parameters as inputs to the network. The spectral parameters obtained were the blackbody temperature, $kT_b$, power-law slope $\Gamma$, flux ratio of the blackbody to power-law components, and reduced $\chi^2$. They also fitted the spectrum, fixing the error on the count rate to be $3\%$, and quoted the reduced $\chi^2$ obtained as a variance, $var$.

\begin{figure*}
    \centering
    \includegraphics[width=0.48\linewidth]{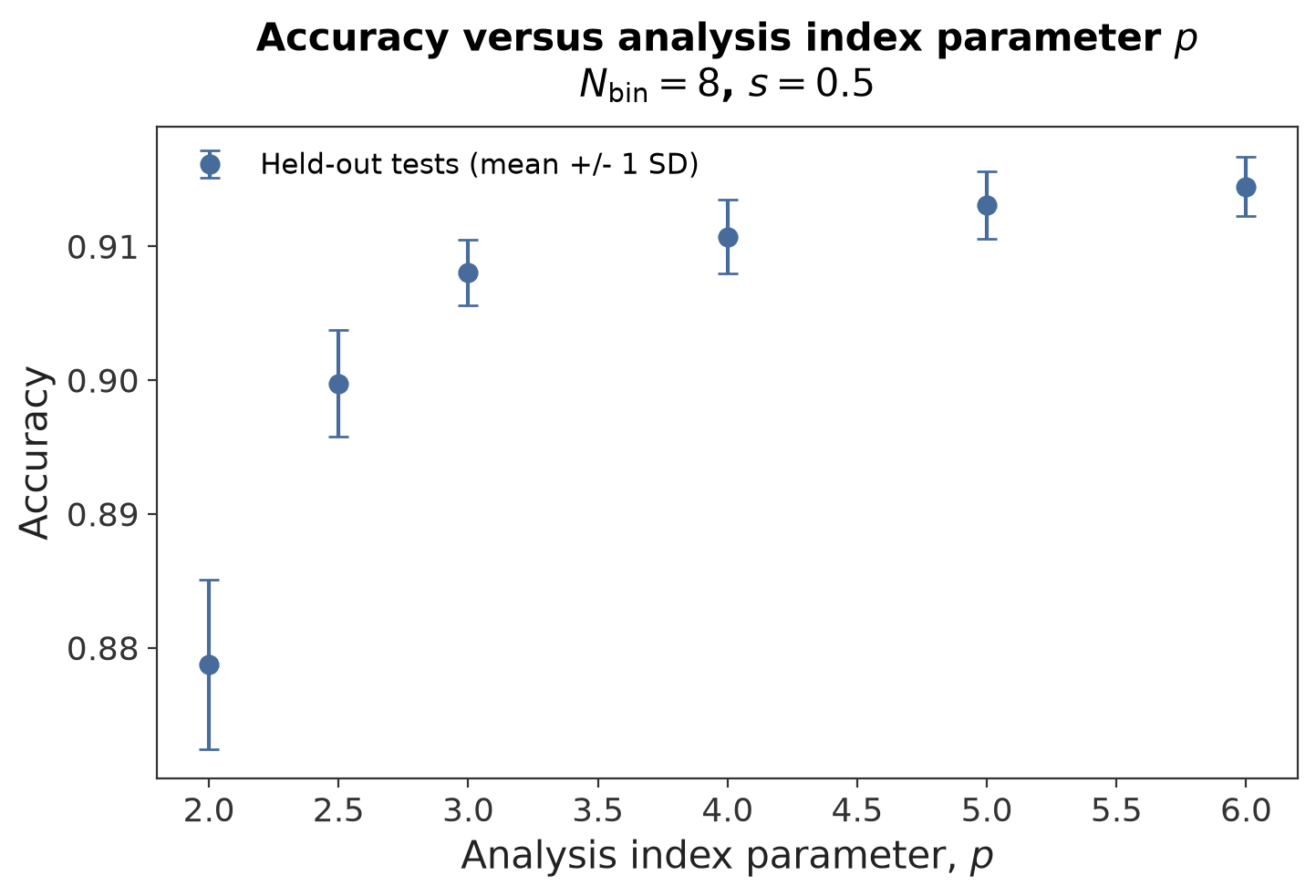}
    \includegraphics[width=0.48\linewidth]{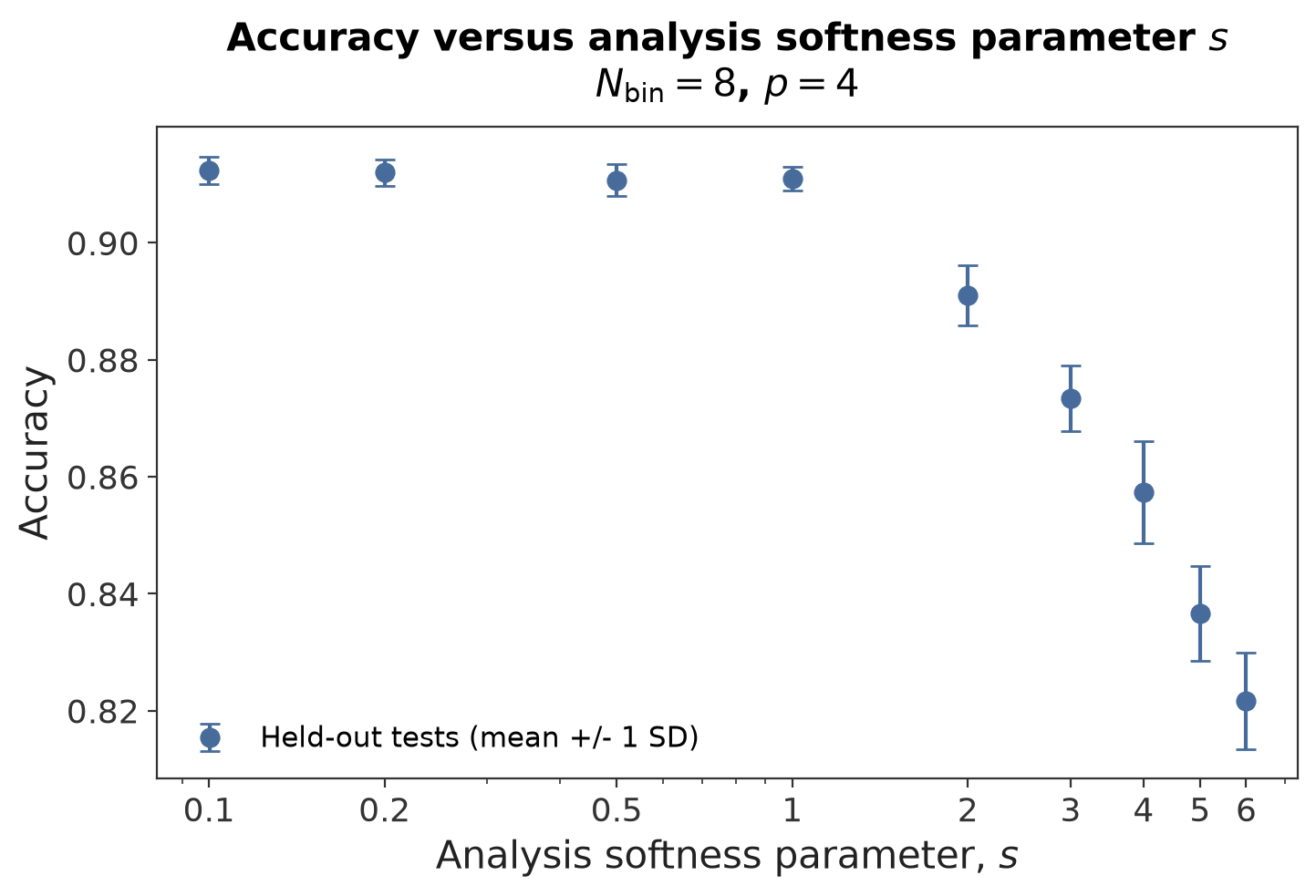}
    \caption{Classification accuracy as a function of the kernel parameters, for $N_{\rm bin}=8$. \textit{Left:} accuracy versus the power-law index $p$, for $s=0.5$. \textit{Right:} accuracy versus the softness parameter $s$, for $p=4$. The mean accuracy over ten random 80\%/20\% splits of the observations into base and test sets is plotted, with the standard deviation as the
    uncertainty. The accuracy saturates for $p>3$ and $s<1$.}
    \label{fig:accuracy_params}
\end{figure*}

In this work, we primarily use these five spectral parameters as the input variables since these allow for an easier interpretation of the results. We have redefined the flux ratio as the blackbody flux to the total flux, $f_{ratio}$, since then the quantity is always less than unity and not extremely large for some observations. Also, there were some slew data in the sample, which we have removed.

\subsection{Classification Accuracy}

To test the accuracy of the classification scheme, the data set is divided into two groups, the base and testing data sets. The base data set is generated by randomly choosing 80\% of all the observations, and the remaining 20\% is kept as the testing data set. The Incidences are computed from the base data set, and the probabilities are computed for each of the observations in the testing data set using Equation \ref{prob}. The accuracy is defined as the fraction of test observations for which the probability of the class is greater than 0.5. The procedure is repeated for ten realisations of the base and test data sets, and the average accuracy and its standard deviation are noted.

Figure \ref{fig:accuracy_bins} (Left panel) shows the average accuracy and its standard deviation as a function of $N_{bin}$, which is an analysis parameter. The other analysis parameters have been fixed at $s = 0.5$ and $p = 4$. For $N_{bin} > 8$ the accuracy saturates at $\sim 0.93$. Note that this is a similar accuracy to that obtained by \cite{garg26} using a Deep Learning technique. Figure \ref{fig:accuracy_params} shows the variation of accuracy for different values of $p$ and $s$. Again, the accuracy saturates for $s < 1$ and for $p > 3$. Thus, the results are not sensitive to the analysis parameters (beyond a certain value), and henceforth we adopt values of $N_{bin} = 8$, $s = 0.5$ and $p = 4$. We have tested that if, instead of a power-law form (i.e., $\propto 1/(D+s)^p$), a Gaussian Kernel (i.e., $ \propto e^ {- (D/h)^2}$, where $h$ is the scaling parameter) is used, similar accuracy is obtained. However, the maximum accuracy is obtained for different values of $h$, which depend on the value of $N_{bin}$. Although there are two parameters for the power-law form, $s$ and $p$, the results are not sensitive to the values used beyond a certain range, which provides an implementation advantage.

For understanding how the analysis is being undertaken, Figure \ref{fig:accuracy_bins} (Left panel) also shows, as a solid line, the accuracy obtained when the entire data set is taken to be both the base and test data sets. Till $N_{bin} < 6$ the accuracy is similar to that obtained from splitting the data, but for larger $N_{bin}$ the accuracy increases to unity. For larger $N_{bin}$, the dimensions become large, and the vectors which have non-zero Incidences become sparse. Hence, the probability computed for a non-zero Incidence vector is determined only by its own incidence value, resulting in all observations being identified correctly. In other words, one can loosely state that the analysis 'remembers' the data set for large $N_{bin}$.

For completeness, the accuracy has also been estimated when the raw count spectra, with all 43 spectral bins, are provided as the data set instead of the five derived spectral parameters. The accuracies as a function of $N_{bin}$ are plotted in Figure~\ref{fig:accuracy_bins} (right panel). We emphasize that this is done for illustrative purposes. The rest of the work is based on the five derived spectral parameters, since it is more straightforward to extract science information from the derived spectral parameters as compared to the raw counts spectra.

%Due to limitations on the number of variables the present analysis code can handle, the raw count spectra were grouped to 22 spectral bins (from the original 42 bins). The accuracies as a function of $N_{bin}$ are plotted in Figure 3.

\subsection{Probability Distribution and most probable class characteristics}

\begin{figure}
    \centering
    \includegraphics[width=\linewidth]{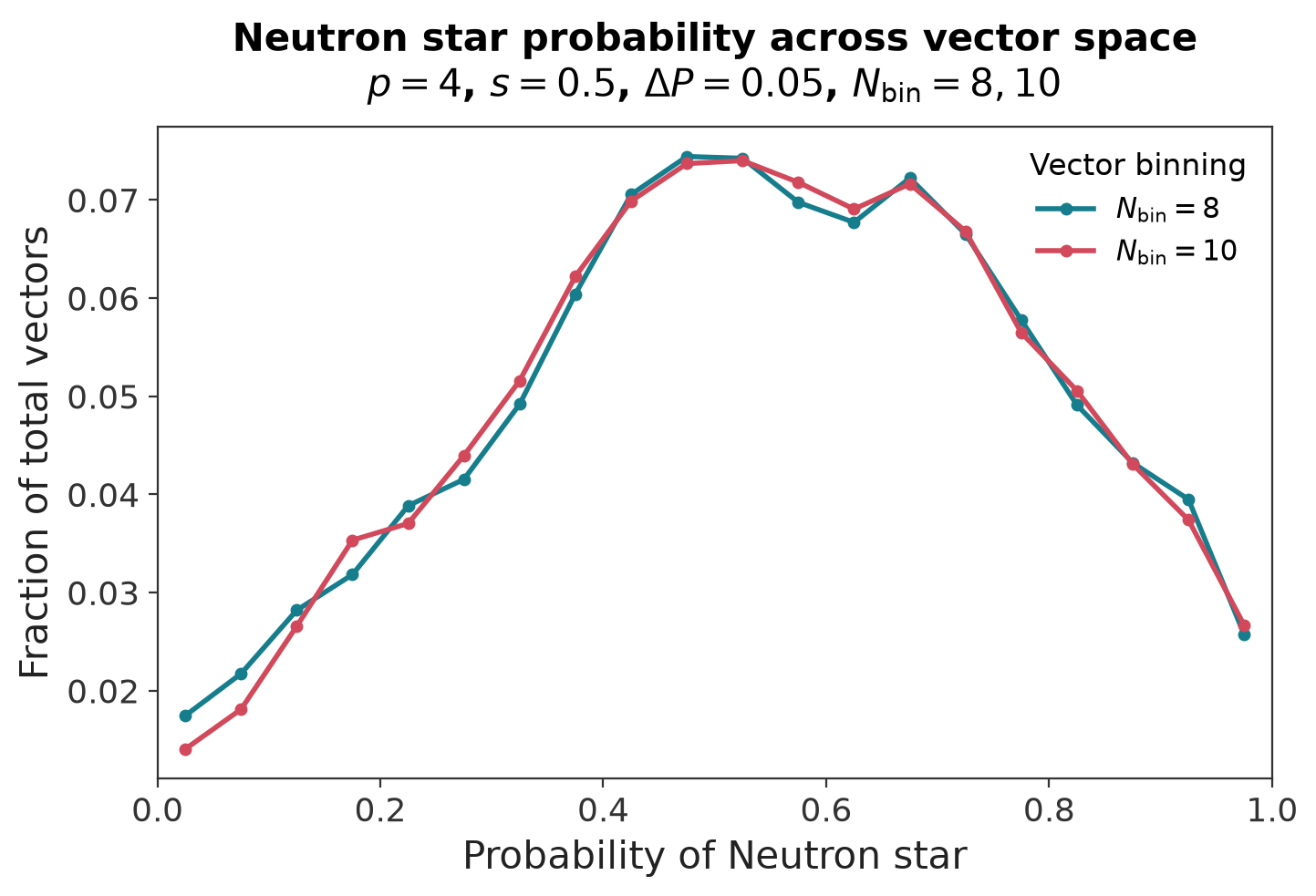}
    \caption{Distribution of the fraction of all possible vectors as a function of the probability $p_{\rm NS}$ of being a neutron star system, for $N_{\rm bin}=8$, $s=0.5$ and $p=4$. The distribution for $N_{\rm bin}=10$ is also shown.}
    \label{fig:prob_dist}
\end{figure}

For  $N_{bin} = 8$, $s = 0.5$ and $p = 4$, the probabilities were computed for all possible vectors. The total number of possible vectors is $N_{bin}^{N_{var}}= 32768$. Figure \ref{fig:prob_dist} shows the distribution of the fraction of the total vectors as a function of the probability for  Neutron star systems. Also plotted is the distribution when $N_{bin} = 10$, which shows that this distribution is roughly independent of the chosen $N_{bin}$. The distribution has a peak at $\sim 0.5$, which implies that for a large fraction of the vectors (i.e., for a large fraction of combinations of possible spectral parameters), it is not possible to distinguish between the two classes. However, most of these vectors do not appear in the data set (i.e., such combinations of spectral parameters are not observed). Hence, the accuracy as shown in the previous section is high. There is also a peak at probability $\sim 0.7$, which implies that if we consider all possible combinations of the spectral parameters, a combination would be more likely to be a neutron star system. In other words, about 60\% of the total vectors have a probability greater than 0.5 of being a neutron star system.

\begin{figure}
    \centering
    \includegraphics[width=\linewidth]{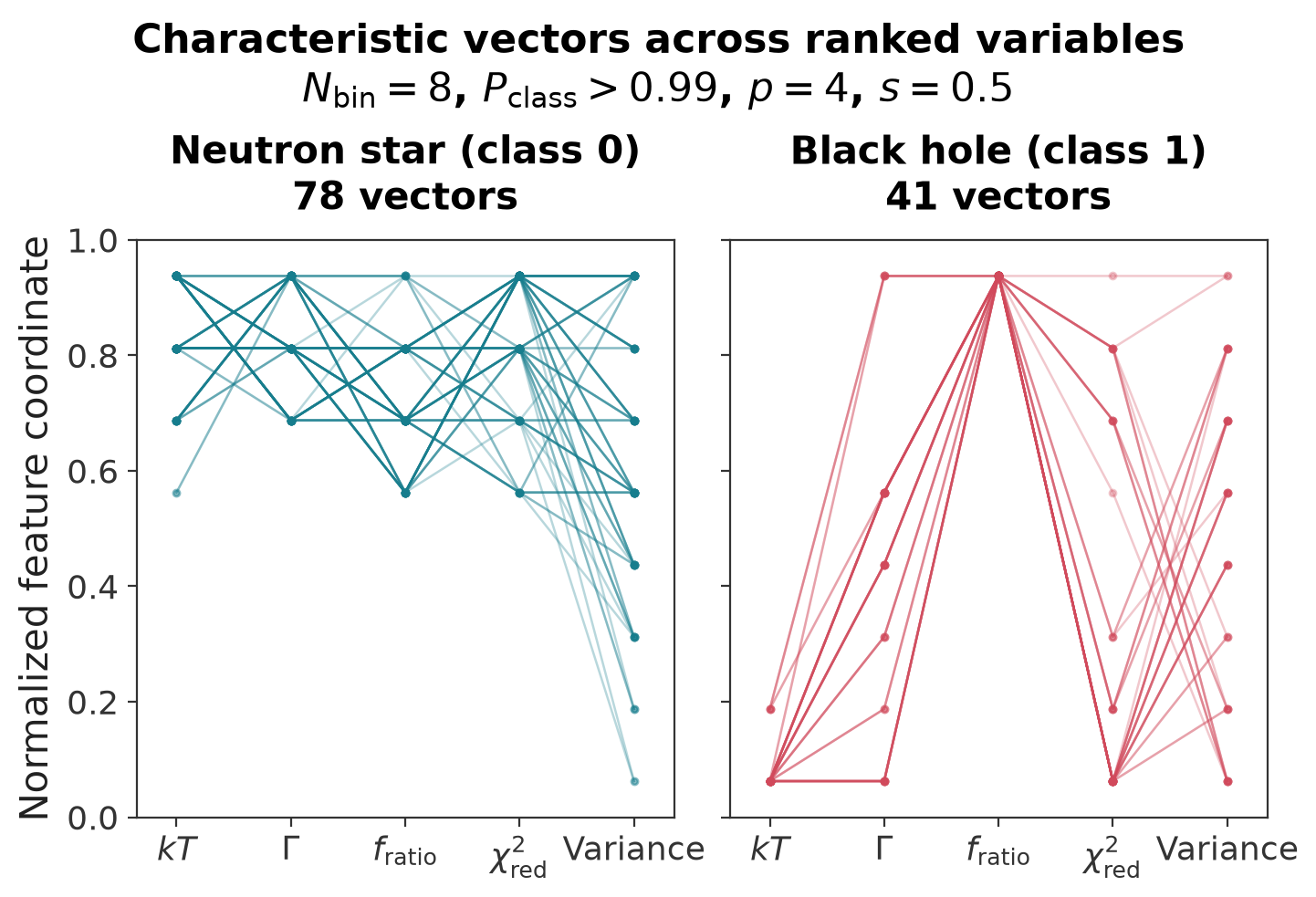}
    \caption{Characteristic vectors with probability $>0.99$ of being a neutron star (\textit{left}, 78 vectors) or a black hole (\textit{right}, 41 vectors), for $N_{\rm bin}=8$, $p=4$ and $s=0.5$. Each vector is drawn as a line joining
    its rank-bin value, normalised to $0$--$1$, for each of the five spectral parameters: blackbody temperature $kT_b$, photon index $\Gamma$, blackbody flux ratio, reduced $\chi^2$ and variance.}
    \label{fig:characteristic_vectors}
\end{figure}

For $N_{bin} = 8$, the number of vectors with probability $> 0.99$ to be a neutron star is $\sim 78$, while that for a black hole is $\sim 41$. Although we have now identified the five-dimensional vectors (or the set of parameters) which the analysis reports as most likely ($> 0.99$) to be a black hole or a neutron star system, it is difficult to represent them in figures. In Figures \ref{fig:characteristic_vectors} (left) and (right), we adopt the scheme where we plot the rank values of these vectors versus the spectral parameters and join the points to show the vectors.  Figure \ref{fig:characteristic_vectors} (right panel) may be interpreted as follows. Observations with blackbody temperature less than a rank value of 0.25 (corresponding to $<$ 1.05 keV) and blackbody flux ratio rank greater than 0.88 (corresponding to $>$ 0.6)  have a high probability of being a black hole system. However, this is not the complete picture, since the other parameters are also playing a role.  Nevertheless, the analysis picks up the canonical soft state spectra of black hole X-ray binaries as being distinct from the Neutron star systems. It is interesting to note that it often requires low values of reduced $\chi^2$, which means that the spectra should be well described by a power-law and blackbody model. Note that since RXTE spectra are for energies $> 3$ keV, blackbody and disc spectra will not be distinguished for temperatures $< 1$ keV.

Figure \ref{fig:characteristic_vectors} (left panel) shows that observations with high blackbody temperature, $kT_b > 1.6$ keV, high photon indices ($\Gamma > 3$), $f_{ratio} > 0.3$, and large reduced $\chi^2 > 1.6$ are likely to be Neutron star systems. Spectra of Neutron star systems are often complex, requiring multiple components such as a blackbody, a disc, and low-temperature Comptonization emission. The analysis is probably identifying this complexity when it ascribes these sets of parameters as being Neutron star systems. Note that systems with low blackbody flux ratio are not present in Figures \ref{fig:characteristic_vectors},  implying that spectra which can be described by a power-law emission are not easily distinguishable as being from either a black hole or a Neutron star system.

It is perhaps important to note that the above sentences are an interpretation of the results presented in Figure \ref{fig:characteristic_vectors}, and it is possible to have alternate interpretations. Moreover, there is more information in the figures for the black hole case; there are combinations of low temperature with the highest spectral index values,  $\Gamma > 3.3 $, which are distinct from the cases when $\Gamma$ is low.

\subsection{Unique Sources}

For the entire data set, the Uniqueness values (Equation \ref{unique}) were computed for all vectors which had non-zero Incidences for $N_{bin} = 8$ and $pd= 3$. The top ten vectors with the highest uniqueness were identified, and then the data set was searched for observations that corresponded to these ten vectors.  There were 103 such observations, of which 96 were of the Neutron star system, 4U 1907+097. \cite{pattnaik} had listed observations of 4U 1907+097 as being from the black hole system, XTE J1908+094. \cite{garg26} noted this mistake and reclassified these observations as being from a neutron star system. However, 4U 1907+097 is a high-mass neutron star binary  {\it pulsar}, which should not have been included in this sample of low-mass X-ray binaries. Fitting a blackbody and power-law component to the highly complex pulsar spectrum results in high values of all five spectral parameters, which the analysis picks up as being unique. Thus, the analysis can identify obvious errors in the data selection process. This also indicates that perhaps, given a combination of spectra from pulsars and non-pulsating sources, the analysis will be able to effectively differentiate between the two.

Next, we compute the Uniqueness using only the black hole system observations. Note that this requires a complete re-analysis, i.e., computation of the incidences again, since the rank order of the variables is different for only the black hole observations as compared to total data sets. We identify the top ten vectors with the highest Uniqueness and find the 28 corresponding observations.  Of these 28  observations, 9 were from GRO J1655-40, and 13 were from XTE J1550-564. These 9 observations of GRO J1655-40 are the same as the ones reported by \cite{uttley15} as belonging to the unusual soft state, which they name as ``hypersoft state''. As in the standard soft state, the spectra of the hypersoft state have a disc-dominated thermal emission, but the high-energy power-law index is significantly steeper than the canonical one. \cite{uttley15} compare the hypersoft spectra to Cyg X-3, which is not part of the sample, being a high-mass X-ray binary. In the literature, the observations of XTE J1550-564 identified by the analysis have not been directly reported as unusual. However, \cite{sobczak2000} have performed an extensive and complete analysis of RXTE observations of this source. From the values given in Figure 4 of their paper, there was an extended period during which the source was in the soft state. In the middle of this extended period, the photon index increased significantly from the canonical value of $\sim 2.5$ to $\sim 4$. This is similar to the hypersoft definition given by \cite{uttley15}. Thus, the analysis identifies the previously reported unusual hyper-soft state of black hole systems, and only two sources (in the data set used here) exhibit this state.

\begin{figure*}
    \centering
    \includegraphics[width=0.48\linewidth]{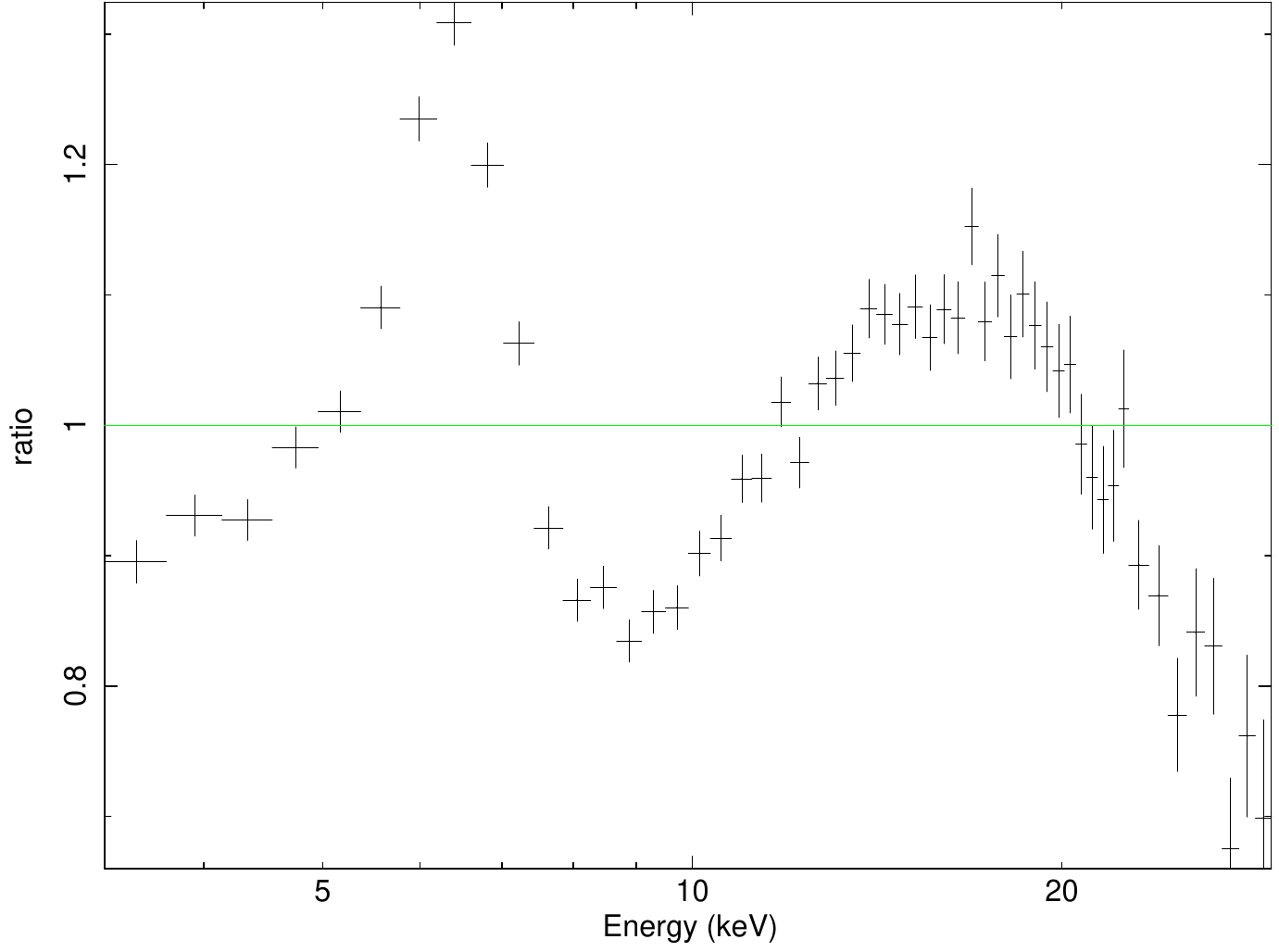}
    \includegraphics[width=0.48\linewidth]{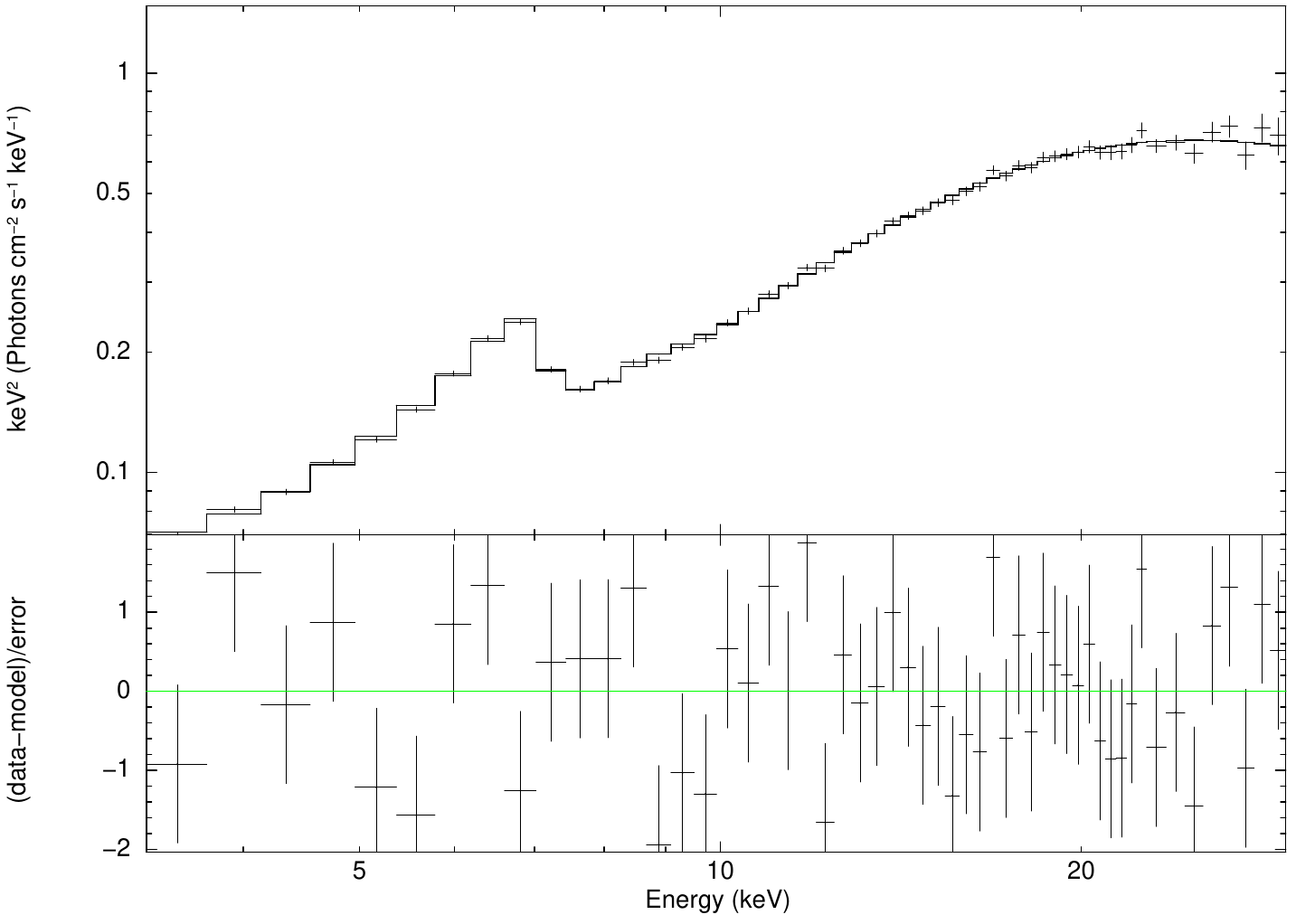}
    \caption{RXTE/PCA spectrum of EXO~1745$-$248 (Obs ID 50054-06-03-01), one of the nine neutron star observations selected as unique. \textit{Left:} ratio of the data to a power-law model, showing a broad iron line and a reflection
    bump. \textit{Right:} the unfolded spectrum fitted with a pure reflection model, {\it relxillCp}, with the residuals, (data$-$model)/error, in the lower panel.}
    \label{fig:exo1745}
\end{figure*}

Next, we redo the analysis only for Neutron Star systems after removing the pulsar 4U 1907+097 from the data set. There were 42 observations corresponding to the ten vectors with the highest Uniqueness, and they were from different sources. An additional criterion was imposed that the reduced $\chi^2 > 1$. This resulted in 9 observations (all of them having reduced $\chi^2 > 15$), which shows that they form a distinct set. Of these 9 observations, 4 are from a single source, EXO 1745-248, which is located in the stellar cluster Terzan 5. In the literature, we did not find anything particular about the spectra of these observations of EXO 1745-248. The left panel of Figure \ref{fig:exo1745} shows the ratio of the spectral counts of one of these observations (Obs ID: 50054-06-03-01) when fitted with a power-law model. There is a clear indication of a broad Iron line and a reflection bump. Indeed, as shown in the right panel of Figure \ref{fig:exo1745}, the spectra can be well described as a pure gravitational-reflection component modelled using the Xspec model {\it relxillCp} \citep{garcia14}. Such reflection-dominated spectra have been reported for Active Galactic Nuclei \citep[e.g.,][]{fabian04,fabian09} but are very unusual for an X-ray binary, and that too for a Neutron star system. A prominent and broad Iron line for  EXO 1745-248 has been reported for an RXTE observation  (Obs ID: 96316-01-40-00) by \cite{iwakiri2021}. This observation is part of the data set, and we find that it corresponds to a vector which is ranked 91st in Uniqueness. The essential difference between this observation and the one shown in Figure \ref{fig:exo1745} is that it does not show a prominent bump at $\sim 20$ keV. A detailed analysis and interpretation of these spectra is beyond the scope of this work and is deferred to a later one.

The choice of examining only the top ten vectors with the highest Uniqueness values and picking up some of the observations as being interesting is somewhat arbitrary. Moreover, one may get slightly different results for different values of the analysis parameters $N_{bin}$ and $pd$. A comprehensive strategy would be to examine a larger number of observations with high Uniqueness values. However, it is also obvious that the observations with the highest Uniqueness may not be interesting, and there would be interesting observations which do not have high Uniqueness. Nevertheless, a scheme like this allows for some distillation of a large data set to some observations which may then be studied in detail.

\section{Discussion}

A scheme based on rank ordering variables and then using KDE has been presented, which allows for class identification as well as obtaining indications of which characteristics of the object allow for definite identification. The scheme can also isolate some of the unique objects in the sample which may be scientifically interesting to study in detail. As an example, the scheme has been used to analyse RXTE spectra of low-mass X-ray binaries.

While there is a possibility of scaling and generalizing the method (as discussed in the appendix), in its present form, the scheme is limited to data in a certain form and of limited size. Thus, at best it may be used as an additional check for some data sets, and advanced Machine learning techniques, which are more versatile and accurate, should be preferred for class identification. Extraction of distinguishing characteristics of a class will naturally depend on the quality, accuracy, and completeness of the data set. Meaningful interpretation of the results will also depend on the nature of variables that have been used in the data set. In this work, the focus has been on derived spectral parameters of the RXTE observations rather than the raw spectra, since the spectral parameters may be easier to interpret. However, the spectral parameters were derived from fitting a phenomenological model consisting of a power-law and blackbody component. More relevant information may have been obtained if a more physical model such as a Comptonization component or with the inclusion of a disc emission or a Gaussian component to represent the Iron line emission. However, many observations would not have the statistics to constrain all parameters of a complex physical model, leading to gaps or missing data. The present scheme would need to be significantly modified to incorporate such data sets.

The scheme can measure a uniqueness parameter which can be used to identify objects in the data set which may be unique and interesting. It should be noted that there are many different and standard schemes to find isolated objects which are also based on distances (e.g., k-NN distance) and density (e.g., DBSCAN). Random Forest may also be used to identify isolated trees.  The scheme described here may be considered as a similar alternative. The uniqueness functionality is integrated into this scheme with the other functions of class and distinguishing characteristic identification, which may provide better insight into the nature of the sources identified as unique. It is important to emphasize that the objects identified may not all be scientifically interesting, but the scheme provides a smaller number of objects that can be subject to more careful analysis, which has the potential for interesting discoveries.
The scheme has to be experimented and tested on different types of data sets, in order to ascertain its general utility and efficiency.

\begin{acknowledgments}
This research has made use of data from the \textit{Rossi X-ray Timing Explorer} (RXTE), obtained from the High Energy Astrophysics Science Archive Research Center (HEASARC), provided by NASA's Goddard Space Flight Center.
We thank the authors of \cite{pattnaik} for sharing the prepared RXTE data set. This research has made use of NASA's Astrophysics Data System Bibliographic Services and of the arXiv preprint server.
\end{acknowledgments}

\facilities{RXTE (PCA)}

\software{XSPEC, Python, NumPy, Matplotlib}

\begin{contribution}
%%This section gives authors the space to recognize author contributions. The text inside this environment is NOT counted towards the total word quanta. At a minimum, manuscripts are expected to include this text:

%All authors contributed equally to the Terra Mater collaboration.
RM conceived the Direct Incidence Method, wrote the analysis code, and drafted the manuscript. AG and AmK prepared the data set and contributed to the manuscript. AK contributed to the discussions and to the manuscript.
%% But authors are expected to provide more specific details, e.g. 
%%
%%SC was responsible for writing and submitting the manuscript.
%%WWM came up with the initial research concept and edited the manuscript.
%%OTS obtained the funding and edited the manuscript.
%%EBF provided the formal analysis and validation. He also edited the manuscript.
%%GEH Supervised the undergraduates, wrote the software and administers the project github and Zenodo repositories.
%%
%% Authors can use the Contributor Role Taxonomy (CRediT) at
%% for ideas on how write a good statement tailored to their needs.

\end{contribution}

%\facilities{HST(STIS), Swift(XRT and UVOT), AAVSO, CTIO:1.3m, CTIO:1.5m, CXO}

%\software{astropy \citep{2013A&A...558A..33A,2018AJ....156..123A,2022ApJ...935..167A},  
%          Cloudy \citep{2013RMxAA..49..137F}, 
%          Source Extractor \citep{1996A&AS..117..393B}
%          }

\section{Data Availability}

The RXTE/PCA observations used in this work are publicly available through NASA's High Energy Astrophysics Science Archive Research Center (HEASARC). The prepared dataset used for the analysis was obtained from the authors of \citet{pattnaik}. Further information regarding the dataset and the analysis products can be obtained from the corresponding author upon reasonable request.

\appendix

\section{Scalability and Generalization}
\label{scale}

In this appendix, we discuss the possibility of scaling and generalizing the method to different sizes and kinds of data. These are expected to be general directions, while the actual implementation may require more specific techniques and experimentation. 

The version of the method described in Section 2 is limited by the number of vectors that can be assigned to an integer value, which is limited to $2^{16}-1$ based on the standard long integer definition. This can be overcome by using functionalities such as the arbitrary-precision integers built into Python or by having one more array $Index_2(k)$ to represent the additional vectors. The data storage requirement is modest since only the non-zero incidence values have to be stored in arrays $Index(k)$ and $IA(c,k)$, where $c$ is the number of classes and $k$ is the number of non-zero incidence vectors, which has to be necessarily less than the number of rows in the data set, $N_{rows}$. If, in some cases, the data set is very large, then one could perhaps consider storing only those vectors which have incidences larger than some value. The computational requirements are also modest, except when one calculates the probabilities for all possible vectors as has been done to generate Figure \ref{fig:prob_dist}, which is a histogram of the number of vectors per probability bin. However, this need not be done when the number of vectors is too large. Note that the highest probability vectors will nearly always be ones with non-zero incidences, and hence these vectors (Figure \ref{fig:characteristic_vectors}) can be obtained by only calculating the probabilities for vectors with non-zero Incidences.

The method has the potential to be generalized to other data formats. One generalisation is if the class consists of continuous real numbers instead of being represented by integers. In this case, just like the variables, the class values may be rank-ordered and divided into bins, $N_{bin}$, where each bin is represented by a number. This transformation will make the data set in the same format as required for the method described in Section 2. An important specific case for this would be when the variables and class values have the same physical meaning. For example, a time series (e.g., a light curve) where the variables are the preceding $n_{var}$ values and then the class (i.e., the value to be predicted) is the value at a certain time. Thus, the method in principle can be used as a generative programme.

There are data sets where a certain variable may consist of a few integer values. The most straightforward way would be to use these integer values directly as indices of the vectors. In this scheme, if a variable consists of real numbers, then find the rank order and identify the value of the variable with the rank order index. If the variable consists of integer values, use these values as the index. The issue with this scheme will be that while the real variables will have  $N_{bin}$ index values, the integer variables will have a different number of index values. Whether to renormalise when computing the distance between vectors or not may require experimentation. Here we have assumed that even when the variables are represented by integers, distance between them still has meaning, i.e., the index 1 is closer to 2 than to 3. For example, this will be the case when the integers represent 'low', 'medium' and 'high'. However, when this is not the case, a distance measure for that variable has to be defined on a case-by-case basis.
There may be data gaps or missing data for a certain variable. If data is missing for a certain variable in the test sample, then it is straightforward not to include that variable in the distance calculation given by Equation \ref{eq2}.
However, it is not clear what needs to be done if a data value is missing in the base set for calculating the incidences. Perhaps one can assign the missing value as the median one or perhaps add a weighted incidence to all values of the variable.  

In general, it will be possible to scale and generalize the method, but this can be done in different ways and experimentation on several different kinds of data sets will be required to determine the optimal method.

\bibliography{sample701}{}
\bibliographystyle{aasjournalv7}



\end{document}